# Document Clustering using Sequential Information Bottleneck Method

[1] MS. P.J.Gayathri

[1] M.Phil scholar, P.S.G.R. Krishnammal College for Women, Coimbatore, India.

[1] gaya3jayaram79@gmail.com

[2] MRS. S.C. Punitha

[2] HOD, Department of Computer science, P.S.G.R. Krishnammal College for Women, Coimbatore, India.

[2] saipunith@yahoo.co.in

[3] Dr.M. Punithavalli

[3] Director , Department of Computer science, Sri Ramakrishna college of Arts and Science for Women, Coimbatore, India.

*Abstract*-**Document clustering is a subset of the larger field of data clustering, which borrows concepts from the fields of information retrieval (IR), natural language processing (NLP), and machine learning (ML). It is a more specific technique for unsupervised document organization, automatic topic extraction and fast information retrieval or filtering. There exist a wide variety of unsupervised clustering algorithms. In this paper presents a sequential algorithm for document clustering based with an enhancement on the features of the existing algorithms. This paper illustrates the Principal Direction Divisive Partitioning (PDDP) algorithm and describes its drawbacks and introduces a combinatorial framework of the Principal Direction Divisive Partitioning (PDDP) algorithm, then describes the simplified version of the EM algorithm called the spherical Gaussian EM (sGEM) algorithm and Information Bottleneck method (IB) is a technique for finding accuracy, complexity and time space. The PDDP algorithm recursively splits the data samples into two sub clusters using the hyper plane normal to the principal direction derived from the covariance matrix, which is the central logic of the algorithm. However, the PDDP algorithm can yield poor results, especially when clusters are not well separated from one another. To improve the quality of the clustering results problem, it is resolved by reallocating new cluster membership using the IB algorithm with different settings. IB Method gives accuracy but time consumption is more. Furthermore, based on the theoretical background of the sGEM algorithm and sequential Information Bottleneck method(sIB), it can be obvious to extend the framework to cover the problem of estimating the number of clusters using the Bayesian Information Criterion. Experimental results are given to show the effectiveness of the proposed algorithm with comparison to the existing algorithm.**

I. INTRODUCTION

The Document clustering has become one of the most active areas of research and the development. One of the challenging problems in document clustering that attempts to discover the set of meaningful groups of documents where those within each group are more closely related to one another than documents assigned to different groups. The resultant document clusters can provide a structure for organizing large bodies of text for efficient browsing [15].

Text clustering referred to as Document clustering is closely related to concept of data clustering.The process of clustering aims to discover natural groupings, and thus present an overview of the classes in a collection of documents. A good clustering can be viewed as one that organizes a collection into groups such that the documents with in each group are both similar to each other and dissimilar to those in other groups. Clustering can either produce disjoint or overlapping partitions. In an overlapping partition, it is possible for a document to appear in multiple clusters. The first challenge in a clustering problem is to determine which features of a document are to be considered discriminatory. A majority of existing clustering approaches choose to represent each document as a vector, therefore reducing a document to a representation suitable for traditional data clustering approaches [18].

There exist a wide variety of unsupervised clustering algorithms that has been intensively studied in the document clustering problem. Among the algorithms that remain the most common and effectual, the iterative optimization clustering algorithms have been demonstrated reasonable performance for document clustering, e.g. the Expectation Maximization (EM) algorithm and its variants, and the well-known K-means algorithm. The K-means algorithm can be considered as a special case of the EM algorithm which has vast vicinity [3] by assuming that each cluster is modeled by a spherical Gaussian, each sample is assigned to a single cluster, and all mixing parameters are equal. The competitive advantage of the EM algorithm is that it is fast, scalable, and easy to implement. Hence, it has been chosen to enhance the algorithm, Expectation - Maximization is proposed, Spherical Gaussian EM algorithm.

Principal Direction Divisive Partitioning algorithm was developed by Boley [1] which is a hierarchal clustering algorithm that performs by recursively splitting the data samples into two sub - clusters. It applies the concept of the Principal Component Analysis for the requirement of the principal eigenvector, which is not computationally expensive. It can also generate a hierarchal binary tree that inherently produces a simple taxonomic ontology. The clustering results produced by the PDDP algorithm compare favorably to other document clustering approaches, such as the agglomerative hierarchal algorithm and associative rule hyper graph clustering. In some cases, the clusters are not well separated from one another, it can yield poor results.

IB Method that performs accuracy solves the complexity but time space is more. Sequential clustering algorithm might find clusters that have high accuracy, and evaluate the algorithm. The results demonstrate the superiority of *sIB* method gives high quality of clusters. In a larger dataset, it reduces the complexity. It gives exact solution for a problem but it needs more time space.

A text clustering based on variant of EM algorithm with a novel algorithm and IB Method with a Sequential algorithm is proposed in this paper. As discussed above, each algorithm has its own strengths and weaknesses. The proposed methodology overcomes the disadvantages of the PDDP algorithm that uses the PCA for analyzing the data and combines it with the EM algorithm as the proposed work. In PDDP splits the data samples into two sub clusters based on the hyper plane normal to the principal direction derived from the covariance matrix of the data. When the principal direction is not representative, the corresponding hyper plane tends to produce individual clusters with wrongly partitioned contents. One practical way to deal with this problem is to run the EM algorithm on the partitioning results. A simplified version of the EM algorithm called the spherical Gaussian EM algorithm is presented for performing such task. IB algorithm both in term of complexity and quality of clusters it finds. Comparing the above-mentioned algorithm yields best result in sIB Method takes more time. Furthermore, based on the theoretical background of the spherical Gaussian EM algorithm and sIB Method naturally extending this framework to cover the problem of estimating the number of clusters using the Bayesian Information Criterion [9].

The paper is organized as follows. Section 2 briefly reviews some important backgrounds of the PDDP algorithm, and addresses the problem causing the incorrect partitioning. Section 3 presents the algorithms like spherical Gaussian EM algorithm and Sequential Information Bottleneck algorithm. Section 4 explains the data sets and the evaluation method, and shows experimental results. Finally, this paper concludes in Section 5 with some directions of future work.

II. DOCUMENT CLUSTERING VIA LINEAR PARTITIONING HYPER PLANES

Considering a one-dimensional data set, e.g. real numbers on a line, the question is how to split this data set into two groups. One simple solution may be the following procedures. The mean value of the data set is first found and then it is compared to each point with the mean value. If the point value is less the mean value, it is assigned to the first group. Otherwise, it is assigned to the second group. The problem arises when it has a dimensional data set. Based on the idea of the PDDP algorithm, this problem can be dealt by projecting all the data points onto the principal direction the principal eigenvector of the covariance matrix of the data set, and then the splitting process can be performed based on this principal direction. In geometric terms, the data points are partitioned into two sub clusters using the hyper plane normal to the principal direction passing through the mean vector [1]. This hyper plane is referred as the linear partitioning hyper plane. Figure 1 illustrates the principal direction and the linear partitioning hyper plane on the 2d2k data set, containing 1000 points distributed in 2 Gaussians.

The PDDP algorithm begins with all the document vectors in a large single cluster. This procedure continues by recursively splitting the cluster into two sub clusters using the linear partitioning hyper plane according to the discriminant functions of the algorithm. This procedure terminates by splitting based on some heuristic, e.g. a pre defined number of clusters. Finally a binary tree is yielded out as the output, whose leaf nodes form the resulting clusters. To keep this binary tree balanced, it selects an unsplit cluster to split by using the scatter value, measuring the average distance from the data points in the cluster to their centroid.

The severe problem of the PDDP algorithm is that it cannot achieve good results when clusters are not well separated from one another. This figure 2 and 3 illustrates this drawback. Figure 2 shows two partitions produced by performing the first iteration of the PDDP algorithm on a 2dimensional data set. The data set consists of 334 points. The actual class labels are not given, but one can observe that it is composed of five compact clusters [8]. Based on the principal direction and the corresponding linear partitioning hyper plane, it can be seen that the PDDP algorithm starts with significantly wrong partitioning on the middle left hand cluster. Figure 3 shows three partitions after the second iteration. If the partitioning is further performed without making some adjustments, the resulting clusters become worse. This indicates that the basic PDDP algorithm can produce poor solutions in some distributions of the data, which cannot be known in advance. Also, it may require some information to suggest whether to split the particular cluster or whether to not split on further.

### III. ALGORITHMS
*A. Estimating Number of document clusters*

The clustering algorithm is applied to a new data set having little knowledge about its contents, fixing a predefined number of clusters is too strict and inefficient to discover the latent cluster structures. The finite mixture model of EM algorithm covers the problem of estimating the number of clusters in the data set. A model selection technique is applied called the Bayesian Information Criterion (BIC) [9]. Generally, the problem of model selection is to choose the best one among a set of candidate models.

The BIC contains two components, where the first term measures how well the parameterized model predicts the data, and the second term penalizes the complexity of the model [4]. Thus, the model selected has the largest value of the BIC,
$M^* = \mathrm{argmax}_i BIC(M_i)$.

As a result, the value is directly obtained of the first term of the BIC from running the sGEM algorithm as well as sIB Method. However, it can also be compute it from the data according to the partitioning. The number of parameters is the sum of k − 1 component probabilities, k · d centroid coordinates, and 1 variance.

Boley's subsequent work [2] also suggests a dynamic threshold called the centroid scatter value (CSV) for estimating the number of clusters. This criterion is based on the distribution of the data. Since the PDDP algorithm is a kind of the divisive hierarchical clustering algorithm, it gradually produces a new cluster by splitting the existing clusters. As the PDDP algorithm proceeds, the clusters get smaller. Thus, the maximum scatter value in any individual cluster also gets smaller. The idea of the CSV is to compute the overall scatter value of the data by treating the collection of centroids as individual data vectors. This stopping test terminates the algorithm when the CSV exceeds the maximum cluster scatter value at any particular point.

The CSV is a value that captures the overall improvement, whereas the BIC can be used to measure the improvement in both the local and global structure. As mentioned earlier, in the splitting process, some information is needed to make the decision whether to split a cluster into two sub clusters or keep its current structure. The BIC is first calculated locally when the algorithm performs the splitting test in the cluster. The BIC is calculated globally to measure the overall structure improvement. If both the local and global BIC scores improve, it is then split the cluster into two children clusters.

*B. Spherical Gaussian EM Algorithm*

It is possible to refine the partitioning results by reallocating new cluster membership. The basic idea of the reallocation method [12] is to start from some initial partitioning of the data set, and then proceed by moving objects from one cluster to another cluster to obtain an improved partitioning. Thus, any iterative optimization-clustering algorithm can be applied to do such operation. The problem is formulated as a finite mixture model, and applies a variant of the EM algorithm for learning the model.

The most critical problem is how to estimate the model parameters. The data samples are assumed to be drawn from the multivariate normal density in $R_d$ also assume that features are statistically independent, and a component $c_j$ generates its members from the spherical Gaussian with the same covariance matrix [5].

```
begin
  Initialization: Set $(z_i)_j^{(0)}$ from a partitioning of the
  data, and $t \leftarrow 0$.
  repeat
    E-step: For each $d_i, 1 \leq i \leq n$ and $c_j, 1 \leq j \leq k$,
    find its new component index as:
    $(z_i)_j^{(t+1)} = \begin{cases} 1, & \text{if } j^* = \text{argmax}_j \log(P^{(t)}(c_j|d_i;\theta_j)) \\ 0, & \text{otherwise.} \end{cases}$

    M-step: Re-estimate the model parameters:
    $P(c_j)^{(t+1)} = \frac{1}{n}\sum_{i=1}^{n}(z_i)_j^{(t+1)}$
    $m_j^{(t+1)} = \sum_{i=1}^{n} d_i (z_i)_j^{(t+1)} / \sum_{i=1}^{n}(z_i)_j^{(t+1)}$
    $\sigma^{2(t+1)} = \frac{1}{n \cdot d}\sum_{i=1}^{n}\sum_{j=1}^{k} \| d_i - m_j \|^2 (z_i)_j^{(t+1)}$.
  until $\Delta \log L_c(\Theta) < \delta$;
end
```

Figure 1   A brief SGEM Algorithm.

Figure 1 gives an outline of a simplified version of the EM algorithm. The algorithm tries to maximize log $L_c$ at very step, and iterates until convergence. For example, the algorithm terminates when $\Delta \log L_c < \delta$, where $\delta$ is a pre defined threshold.

### C. Information Bottleneck Method

Clustering algorithms start either from pair wise 'distances' between points pair wise clustering or with a distortion measure between a data point and a class centroid. In the context of document clustering, measure of similarity of two documents is the similarity between their word conditional distributions. Specifically, for every document we define

$$p(y|x) = \frac{n(y|x)}{\sum_{y' \in Y} n(y'|x)}$$  ---1

Where $n(y|x)$ is the number of occurrences of the word $y$ in the document $x$. To avoid an undesirable bias due to different document lengths we also require a uniform prior distribution, $p(x) = 1/|X|$, where $|X|$ is the number of documents in the corpus. This formulation of finding a cluster hierarchy of the members of one set (e.g., documents), based on the similarity of their conditional distributions with respect to the members of another set (e.g., words), was first introduced in [19] and was termed "distributional clustering".

The issue of selecting and justifying the 'right' distance measure between distributions remains, however, unresolved. A principled approach to this problem, which avoids the arbitrary choice of a distortion measure. In this approach, given the joint distribution $p(X, Y)$, which $X$ preserves as much information as possible about the relevant variable $Y$. The mutual information, $I(X; Y)$, between the random variables $X$ and $Y$ is given by (e.g., [2]).

$$I(X;Y) = \sum_{x \in X, y \in Y} p(x)p(y|x) \log \frac{p(y|x)}{p(y)}$$  ---2

and is the natural statistical measure of the information that variable $X$ contains about variable $Y$. It is argued that both the compactness of the representation and the preserved relevant information are naturally measured by mutual information. We introduce a compressed representation $T$ of $X$, by defining $P(T | X)$. The compactness of the representation is now determined by $I(T;X)$, while the quality of the clusters, $T$, is measured by the fraction of the information they capture about $Y$, namely, $I(T;Y) = I(X; Y)$. This general problem has an exact optimal formal solution without any assumption about the origin of the joint distribution $p(X, Y)$ This solution is given in terms of the three distributions that characterize every cluster $t \in T$: the prior probability for this cluster, $p(t)$, its membership probabilities $p(t|x)$, and its distribution over the relevance variable, $p(y|t)$. The information bottleneck principle determines the distortion measure between the points $x$ and $t$ to be the $D_{KL}$ $(p(y|x)\|p(y|t)) = \sum_y p(y|x) \log p(y|x) / p(y|t)$, the Kullback-Leibler divergence between the conditional distributions $p(y|x)$ and $p(y|t)$. Specifically, the formal solution is given by the following equations which must be solved self-consistently,

$$\begin{cases} p(t|x) = \frac{p(t)}{Z(\beta,x)} \exp(-\beta D_{KL}(p(y|x)\|p(y|t))) \\ p(y|t) = \frac{1}{p(t)} \sum_x p(t|x)p(x)p(y|x) \\ p(t) = \sum_x p(t|x)p(x), \end{cases}$$  ---3

where $Z(\beta, x)$ is a normalization factor, and the single positive (Lagrange multiplier) parameter $\beta$ determines the trade off between compression and precision and the "softness" of the classification. In this procedure the information contained in $X$ about $Y$ is `squeezed' through a compact `bottleneck' of clusters $T$, which is forced to represent the `relevant' part in $X$ with respect to $Y$.

Consider the following general scenario. We are given a set of objects $X$ and we would like to find a partition $T(X)$ which maximize score function $F(T)$. In this approach we start with a partition of $X$ into singletons, and at each step we greedily choose the merger of two clusters that maximizes the score. We repeat such some greedy agglomeration steps until we get the desired number of clusters, which we will denote by $K$. Agglomerative clustering is particularly attractive when the score function $F$ is *decomposable*, i.e., if $T = \{ t_1..., t_k \}$, then $F(T) = \sum_i F(\{ t_i \})$. In this case, the change in the total score by merging two clusters is simply

$d_F(t_i, t_j) = F(\{t_i\} \cup \{t_j\}) - F(\{ t_i \}) - F(\{t_j\})$.

There are two main obstacles to agglomerative clustering. First, this greedy approach is not guaranteed to find the optimal partition of $X$ into $K$ clusters. Second, an agglomeration procedure has the time complexity.

We describe a simple idea for solving these two problems by casting an agglomerative algorithm into a new sequential clustering procedure. Unlike agglomerative clustering, this procedure maintains a partition with exactly $K$ clusters. We start from an initial random partition $T = \{t1, t2.... t_K\}$ of $X$. At each step, we "draw" some $x \in X$ out of its current cluster $t(x)$ and represent it as a new singleton cluster. Using a greedy agglomeration step we can now merge $x$ into $t^{new}$ such that $t^{new} = \arg \min\ t \in T\ d_F(\{x\}, t)$, to obtain a new partition $T^{new}$ (with the appropriate cardinality). Assuming that $t_{new} \neq t$ it is easy to verify that $F(T^{new}) > F(T)$. Therefore, each such step either improves the score, or leaves the current partition unchanged. If $F(T)$ is known to be upper bounded we are guaranteed to converge to a local maximum in the sense that no more assignment changes can be performed.

Since this algorithm can get trapped in a local optima, we repeat the above procedure for random initializations of $T$ to obtain $n$ different solutions, from which we choose the one which maximize $F(T)$. Finally, to avoid too slow convergence we define two "convergence" parameters denoted by ε and *max L*. Specifically we declare that the algorithm converged if we already performed *max L* loops over $X$ or if in the last loop we got less than ε $|X|$ assignment changes.

```
Input:
    |X| objects to be clustered
    Parameters: K, n, maxL, ε
Output:
    A partition T of X into K clusters
Main Loop:
    For i = 1, ..., n
        T_i ← random partition of X.
        c ← 0, C ← 0, done = FALSE
        While not done
            For j = 1, ..., |X|
                draw x_j out of t(x_j)
                t^{new}(x_j) = arg min_{t'} d_F({x_j}, t')
                If t^{new}(x_j) ≠ t(x_j) then c ← c + 1
                Merge x_j into t^{new}(x_j)
            C ← C + 1
            if C ≥ maxL or c ≤ ε · |X| then
                done ← TRUE
    T ← arg max_{T_i} F(T_i)
```

Figure 2: A Pseudo-code for the algorithm.

In this complexity of sequential approach is in each "drawing" step we should calculate $d_F(\{ x \}, t)$ for every $t \in T$ which is an order of $O(K|Y|)$. Our time complexity is thus bounded by $O(nLK|X||Y|)$ where $L$ is the number of loops we should perform (over $X$) until convergence is attained. Since typically $nLK \ll |X|^2$ we get a run time improvement.

Our significant sequential clustering algorithm is reminiscent of the standard $K$-means algorithm. The main difference, is that $K$-means perform *parallel updates*, in which first we choose for each $x$ its new cluster, and then we move *all* the elements to their new clusters in one step. As a consequence, the definition of the clusters (i.e., their centroids in $K$-means) changes only after all the elements move to their preferred clusters. To show that such a step is justified, we have to require more structure of the target function $F$. We also note here that our sequential framework has some relations to the *incremental* variant of the EM algorithm for maximum likelihood [24], which still needs to be explored.

*a) Sequential IB Clustering*

The application of the above discussion in the context of the *Information Bottleneck* method is straightforward. We define $F(T) = I(T; Y)$ and represent each $x$ by $p(x, y)$. The greedy merging criterion is known from the *Agglomerative Information Bottleneck (AIB)* algorithm [20,22]. Specifically, in this context we get

$$d(x, t) = (p(x) + p(t)) \cdot JS(p(y|x), p(y|t)) ,$$

-- 4

where *JS*(*p, q*) is the *Jensen-Shannon* divergence [21, 23] defined as

$$JS(p,q) = \pi_1 D_{KL}(p\|\bar{p}) + \pi_2 D_{KL}(q\|\bar{p}) ,$$

where in our context

$$\{p,q\} \equiv \{p(y|x), p(y|t)\}$$

$$\{\pi_1, \pi_2\} \equiv \{\frac{p(x)}{p(x)+p(t)}, \frac{p(t)}{p(x)+p(t)}\}$$

$$\bar{p} = \pi_1 p(y|x) + \pi_2 p(y|t) .$$
--5

Any given partition T defines some membership ("hard") probability p (t|x), which in turn defines p(y|t) and p(t) for every t ε T Additionally since I(T; Y ) is indeed upper bounded we are guaranteed to converge to a local maximum of the information.

The JS divergence is non-negative and is equal to zero if and only if both its arguments are identical. It is upper bounded and symmetric, though it is not a metric. The JS-divergence relates it to the (logarithmic) measure of the likelihood that the two sample distributions originate by the most likely common source, denoted here by p [23]. Using this interpretation we can interpret the new algorithm as follows. At each step we draw some x and merge it back into its most probable source. We refer to this algorithm as the 'sIB' algorithm.

## IV. RESULTS AND DISCUSSIONS

### A. Data sets and setup Information

The 20 Newsgroups data set consists of 20000 articles evenly divided among 20 different discussion groups [10]. This data set is collected from UseNet postings over a period of several months. Many categories fall into confusable clusters. For example, five of them are computer discussion groups, and three of them discuss religion. The Bow toolkit [11] is used to construct the term document matrix (sparse format). The UseNet headers are used, and also eliminated the stop words and low frequency words (occurring less than 2 times). Finally 59965×19950 term document matrix is obtained for this data set.

The well known tf·idf term weighting technique is also applied. Let $d_i = (w_{i1}, w_{i2}, w_{im})^T$, where m is the total number of the unique terms. The tf·idf score of each $w_{ik}$ can be computed by the following formula:

$$w_{ik} = tf_{ik} \cdot \log(n/d_{fk})$$

where $tf_{ik}$ is the term frequency of $w_k$ in $d_i$, n is the total number of documents in the corpus, and $df_k$ is the number of documents that $w_k$ occurs. Finally, each document vector is normalized using the $L_2$ norm. For the purpose of comparison, the basic PDDP algorithm is chosen as the baseline. The number of clusters k is varied in the range [2, 2k], and no stopping criterion was used.

### B. Evaluation Method

Since all the documents are already categorized, comparing clustering results with the true class labels can perform evaluation. In our experiments, the normalized mutual information (NMI) is been used [16]. In the context of document clustering, mutual information can be used as a symmetric measure for quantifying the degree of relatedness between the generated clusters and the actual categories. Particularly, when the number of clusters differs from the actual number of categories, mutual information is very useful without a bias towards smaller clusters, by

TABLE1. CLUSTERING RESULTS BY VARYING STOPPING CRITERIA ON 20 NEWSGRIUPS DARA SETS

| Data set | Criterion | Algorithm | k found | NMI | Time (sec.) |
|---|---|---|---|---|---|
| 20 Newsgroups | CSV | PDDP | 34 | 0.443 | 15.838 |
| | | sGEM | 34 | 0.482 | 105.39 |
| | | IB | 34 | 0.524 | 167.42 |
| | BIC | PDDP | 25 | 0.426 | 14.70 |
| | | sGEM | 25 | 0.463 | 78.45 |
| | | IB | 25 | 0.506 | 126.36 |

Normalizing this criterion to take values between 0 and 1, the NMI can be calculated as follows

$$NMI = \frac{\sum_{h,l} n_{h,l} \log(n \cdot n_{h,l}/n_h n_l)}{\sqrt{(\sum_h n_h \log(n_h/n))(\sum_l n_l \log(n_l/n))}}$$

where $n_h$ is the number of documents in the category h, $n_l$ is the number of documents in the cluster l**,** and $n_{h,l}$ is the number of documents in the category h as well as in the cluster l. The NMI value is 1 when clustering results exactly match the true class labels, and close to 0 for a random partitioning [17].

### C. Experimental Results

In this data set, it can be seen that the proposed algorithm perform relatively better than the basic PDDP algorithm. However, performing the global refinement after the local refinement as in EM degrades the quality of the clustering results. The global refinement with the sIB algorithm leads to more decisions to move each document from its cluster to other candidate clusters. IB method yields more accuracy than sGEM algorithm.

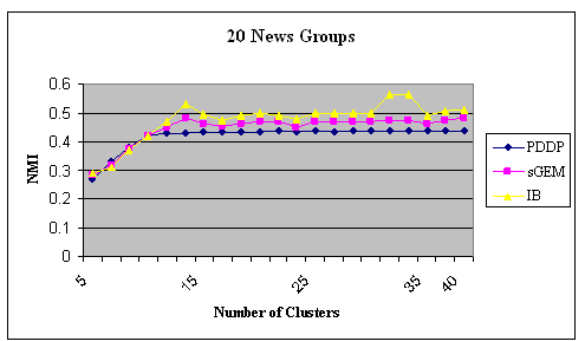

Figure 3: NMI results on the 20 Newsgroups data set.

Further results are discussed of estimating the number of underlying clusters. Note that both the data sets have the actual number of categories at 20. Table 1 summarizes clustering results by varying the stopping criteria. For the 20 Newsgroups data set, with the CSV, PDDP, sGEM and sIB Mehtod find 34 clusters.

Since the BIC measures how well the partitioning can model a given data set, without the local refinement, sGEM decides to keep almost all documents of the Entertainment subcategories in a large single cluster. Although sGEM further splits the data set, a large cluster containing documents of the Entertainment subcategories is still kept. Note that the other main topics can be partitioned into their own clusters, e.g. Health (H) in C2, Sports (S) in C1, Politics (P) in C3, and Business (B) in C5. Large clusters are now partitioned into small sub clusters, since the selection method is just based on the largest scatter value. However, a small number of documents of each Entertainment subcategories remain in the cluster C13. It can be seen that both the BIC and the CSV can yield a reasonable number of clusters, but the contents of clusters are generated different due to their theoretical background concepts.

V. CONCLUSION AND FUTURE WORK

This paper presents several strategies for improving the basic PDDP algorithm. When the principal direction is not representative, the corresponding hyper plane tends to produce individual clusters with wrongly partitioned contents. By formulating the problem with the finite mixture model. This paper describes the sGEM algorithm has tremendous improvement when compared to the PDDP algorithm in several ways for refining the partitioning results. sIB algorithm avoids the complexity of partitioning the clusters than sGEM algorithm. Though its accuracy is high, it takes more time consumptions. Preliminarily experimental results on two different document sets are very encouraging.

In future work, intends to investigate other model selection techniques for approximating the number of underlying clusters. IB Method can be applied in Machine Learning techniques. The statistical measure can also be applied for this algorithm in further enhancement.

## AUTHORS PROFILE

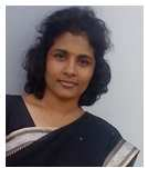

**P.J.Gayathri** is pursuing her M.Phil., in P.S.G.R Krishnammal College for Women, Coimbatore. She completed her M.sc., from Bharathidsan University in the field of Computer Science.Her area of research in Data mining.

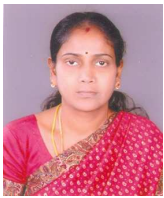

**S.C.Punitha**, HOD, Department of Computer Science and Information Technology, P.S.G.R Krishnammal College for Women, Coimbatore. She completed her M.Sc.,MFT.,M.Phil., from Bharathiar University. She is pursuing her Ph.D in Karunya University, Coimbatore in the field of Computer Science. She has an academic experience of 12 years. Her area of research includes Data mining and Artificial Intelligence.

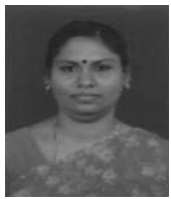

**Dr. M. Punithavalli** received the Ph.D degree in Computer Science from Alagappa University, Karaikudi in May 2007. She is currently serving as the Director of the Computer Science Department, Sri Ramakrishna college of Arts and Science for Women, Coimbatore. Her research interest lies in the area of Data mining, Genetic Algorithms and Image Processing. She has published more than 10 Technical papers in International, National Journals and conferences. She is Board of studies member various universities and colleges. She is also reviewer in International Journals. She has given many guest lecturers and acted as chairperson in conference. Currently 10 students are doing Ph.D under her supervision